\documentclass[10pt]{article}

\usepackage{color}
\usepackage{graphicx}
\usepackage{amsmath,amssymb}
\usepackage{multirow}

\usepackage[numbers]{natbib}
\begin{document}

%\institution{\orgdiv{Department of Methods and Models for Economics, Territory and Finance}, \orgname{Sapienza University of Rome}, \orgaddress{\state{Rome}, \country{Italy}}}

{
  \title{\bf Bayesian inference for discretely observed continuous time multi-state models}
  \author{Rosario Barone \hspace{.2cm}\\
    Sapienza University of Rome, Italy\\
    and \\
    Andrea Tancredi \\
    Sapienza University of Rome, Italy}
  \date{}
  \maketitle
}

\abstract{Multi-state models are frequently applied for representing processes evolving through a discrete set of state.  Important classes of multi-state models arise when transitions between states may depend on the time since entry into the current state or on the time  elapsed from  the starting of the process. The former models are called   semi-Markov while the latter are known as inhomogeneous Markov models. 
Inference for both the models  presents  computational difficulties when the process is only observed at discrete time points with no additional information about the state transitions.  In fact, in both  the cases, the likelihood function is not  available in closed form.   In order to obtain  Bayesian inference under these two classes of models we reconstruct the whole unobserved trajectories conditioned on the observed points via a Metropolis-Hastings algorithm. As proposal density we use that given by the nested Markov models whose conditioned trajectories can be easily drawn by the uniformization technique. The resulting inference is illustrated via simulation studies and the analysis of two benchmark data sets for multi state models.}

\vspace{0.3cm}

{\it keywords}: Metropolis-Hastings, inhomogeneous Markov models, panel data, semi-Markov models

\section{Introduction}\label{sec1}

Continuous-time  multi-state models (CTMSM) are continuous time processes with discrete and finite state space. They represent a useful  class of stochastic processes  for analyzing  event history data \citep{lawless2013armitage}. Practical applications of these  models  can be found in many fields.  For example in biostatistics  they are often used  for modelling both disease progression and patient recovery after medical treatment, see for example Gentleman et al \cite{gentleman1994multi} or Ieva et al \cite{ieva2017multi} for a more recent utilization  Other applications can be found in  econometrics where, for example,  CTMSM  have ben adopted for modeling individual labor market status or credit rating transitions \citep{joutard2012continuous},\cite{bladt2009efficient}.

Inference details for modeling data generated by specific classes of CTMSM differ if the the sample paths are continuously observed or if the data only consist of the states observed  at discrete time  points,  with  no  information  about  the  state sequence   and the jump  times. In fact, the former case does not present particular issues while the latter, usually referred to as panel data  framework,  apart from specific models may present troublesome computational issues.

The problem with the discretely observed  framework is that the likelihood function is available only for the Markov case or its simpler extensions and also in these cases it must be evaluated via numerical approximations. Kalbfleisch and Lawless \cite{lawless85} were the first to introduce appropriate numerical techniques for the standard Markov CTMSM.  Relaxing the Markov assumptions, for example by assuming a semi-Markov process where  transitions between states may depend on the time since entry into the current state, leads generally  to an intractable likelihood problem. To bypass the problem, Kang and Lagaskos \cite{kang2007statistical}  assume time-homogeneous transition intensities from at least one of the states while Armero et al \cite{armero2012bayesian}  discuss a Weibull progressive disability model. Titman and Sharples \cite{titman2010semi} focus on the tractable class of phase-type sojourn distributions while Titman \cite{titman2014estimating} suggests using phase-type distributions to approximate the likelihood of CTMSM with Gamma or Weibull sojourn time distributions.

Another problematic case is that of  the Markov CTMSM with transition rates  depending  on the time elapsed from the starting of the process; i.e. the  time-inhomogeneous Markov processes.  In this class of models,
the Kolmogorov Forward Equations (KFE) providing the transition probabilities do not have a general 
analytical solution although there are   specific cases where analytic solutions can be maintained. For example Kay \cite{kay1986markov} suggests to use piecewise constant transition intensities, while Hubbard et al \cite{hubbard2008modeling} propose models in which the time scale of a nonhomogeneous Markov process is transformed to an operational time scale on which the process is homogeneous, developing a method for jointly estimating the time transformation and the transition intensity matrix for the time transformed homogeneous process. Moreover, numerical quadrature techniques have been used to calculate the transition probabilities for progressive models by P\'erez-Oc\'on et al \cite{perez2001non} and Hsieh et al \cite{hsieh2002assessing}. Finally, a more general approach has been proposed by Titman \cite{titman2011flexible} and 
Machado et al \cite{machado2021penalised} where methods for numerically solving nonlinear differential equations  have been used to estimate inhomogeneous model with semi-parametric transition rates based on spline functions.

All the proposals  discussed above do not  consider the possibility to reconstruct the whole sample paths  in order   to make inference via ordinary missing data techniques.  However,  note that a missing data formulation  
has been frequently adopted in the Markov case starting from Baldt and S{\o}rensen \cite{bladt05}, where both an EM and  a Gibbs sampler were proposed to estimate the parameters of  a discretely observed Markov CTMSM.   Anyway  the Gibbs sampler was performed by a naive rejection sampling, i.e. by drawing unconditioned trajectories and discarding them if they do not hit the right states.  More recently, Luo et al \cite{luo2020bayesian} employed a modified rejection sampling in a Bayesian analysis of a hidden  time inhomogeneous Markov model  by changing the  rate matrix of the proposal Markov process  accordingly to the sub-interval  end-points. 
The limitations of the  rejection sampling were discussed in Hobolth and Stone \cite{hobolth09}, where it was also introduced a different sampling strategy  based on the uniformization technique that permits to simulate directly Markov trajectories with fixed starting and ending states. The uniformization algorithm was used also to draw the distribution of a Markov sample path conditionally on a sequence of observed points by Fearnhead and Sherlock \cite{fearnhead2006exact} and was also used by Pfeuffer et al \cite{pfeuffer2018} for implementing a  stochastic version  of  the EM for the Markov case.  Note that a stochastic EM algorithm was recently proposed also for the semi-Markov case by Aralis and Brookmeyer \cite{aralis2019stochastic}, but the reconstruction of the sample paths was performed by a naive rejection sampling. Finally Tancredi \cite{tancredi2019biom} proposes approximate Bayesian computation (ABC) techniques for Markov and Semi-Markov cases by approximately matching the observed and simulated  state transition matrices between different observation times.

The purpose of this article is to demonstrate that Markov Chain Monte Carlo methods can be efficiently used  to estimate discretely observed multi-state models.  In particular we propose a general MCMC algorithm to obtain
Bayesian inference both for semi-Markov and time-inhoomgeneous Markov. By a Gibbs sampler approach, the algorithm reconstruct the missing structure of the partially observed paths. The simulation of the conditioned semi-Markov or time-Inhomogeneous trajectories is performed by a Metropolis-Hasting step with proposal density provided by the distribution of the conditioned trajectories assuming the Markov model. The uniformization algorithm will be used for an efficient simulation of the conditioned trajectories. Note also that the idea to exploit proposal distributions baed on simpler Markov assumptions for dicrete time series models was used also by Capp\'e et al \cite{cappe2004population} to estimate a discrete hidden semi-Markov model.

The paper is organized as follows. In Section 2 we formally define the continuous time multi-state Markov models and the  semi-Markov and time-inhomogeneous extensions.  In Section 3 we present the uniformization algorithm for the simulation of end-point conditioned Markov CTMSM and  we outline the uniformization based Metropolis-Hasting algorithm for simulating conditioned semi-Markov and time-inhomogeneous Markov trajectories. We provide also the details of  the complete  MCMC algorithm for Bayesian inference. In particular we consider the   Weibull distribution  for the sojourn times in the semi-Markov case and the Gompertz model for the transition  rates  in the inhomogeneous case.
% but our approach can be used with other parametric models. 
In  Section 4 we analyze the results obtained both on simulated and real data sets and in Section 5 we provide a short discussion.

\section{Continuous time multi-state models}

Let  $Y(\cdot)=\{Y(t),t\geq 0\}$ be a continuous time process with state space $\mathcal{S}=\{1,\ldots S\}$.
A CTMSM can be defined via the transition intensity fuction
$$q_{rs}(t,\mathcal{F}_t)=\lim_{\delta t\rightarrow 0}
\frac{
P\{Y(t+\delta t)=s|Y(t)=r,\mathcal{F}_t\}}{\delta t}$$
representing the instantaneous probability of a transition from state $r$ to state $s$ at time $t$ when $\mathcal{F}_t$ is the past history up to time $t$.

Considering
\begin{equation}
\label{markov}
P\{Y(t+\delta t)=s|Y(t)=r,\mathcal{F}_t\}=\left\{
\begin{array}{l l }
\gamma_{rs} \delta t +o(\delta t) & s\neq r\\ 1+\gamma_{rr}\delta
t+o(\delta t) & s=r
\end{array} \right.
\end{equation}
 where $\gamma_{rs}\geq 0$ and $\gamma_{rr}=-\sum_{s\neq r}\gamma_{rs}=-\gamma_{r}$, we have Markov CTMSM.  In (\ref{markov}) the transition probabilities do not depend on $t$, so the process is time-homogeneous.
Note  that $\gamma_{rs}>0$ is the rate at which transitions from $r$ to $s$
occur while $\gamma_r$ is the rate of a transition out of the state $r$. Note also that if $r$ is an absorbing state then $\gamma_{rs}=n_{rs}=0$ $\forall s\neq r$. A formula for the transition probabilities for a Markov CTMSM can be obtained with the Chapman-Kolmogorov forward equations.  Let ${\bf G}$ be the rate matrix 
$$G=\left(
\begin{array}{c c c }
\gamma_{11} & \ldots & \gamma_{1S} \\
\vdots & \ddots & \vdots\\
\gamma_{S1} & \ldots & \gamma_{SS}
\end{array}
\right).$$ 
Then $p_{rs}(t;{\bf G})=$ $P\{Y(u+t)=s|Y(u)=r\}$ is the $(r,s)$ element of the exponential matrix 
$$ \exp{(t{\bf G})}=\sum_{r=0}^{\infty}\frac{t^r}{r!} {\bf G}^r. $$

An equivalent representation for $Y(\cdot)$  can be obtained by considering the  sequence of  visited states  and  sojourn times. In fact, 
let $\{S_i, i\geq 0\}$  be the random sequence indicating  the states visited by the process, let  $\{Z_i, i\geq 1\}$ be the random sequence representing the  times at which the jumps occur and let $\{W_i=Z_{i}-Z_{i-1} , i\geq 1 \}$ be the random sequence with the sojourn times. The process $Y(\cdot)$ is equivalent to the sequence 
$S_0,W_1,S_1,\ldots, W_i,S_i, \ldots \, \,. $
In the homogeneous case  the state sequence $\{S_i, i\geq 0\}$  is a Markov chain with transition probabilities $p_{rs}=\gamma_{rs}/\gamma_r$, $s\neq r$ and the holding times $\{ W_i, i\geq 1\} $ are independent exponential random variables with rates $\gamma_{s_{i-1}}$ depending on the departure state.

Let $y=y(t)$ be a completely  observed trajectory of the process  on $t\in [0,T]$ . Throughout the paper we assume that the initial  state $s_0$ is fixed.  Moreover let $s_1,\ldots,s_n$ be the state sequence in $[0,T]$  and  $z_1,\ldots, z_n$ be the jump times sequence. { For censored trajectories $T$ represents the censoring time}; the density of $y$ can be written as
\begin{equation}
\label{markov2}
p_M(y)=p_M(s,z)=\left( \prod_{i=1}^n p_{s_{i-1} s_i} \gamma_{s_{i-1}} e^{-\gamma_{s_{i-1}}(z_i-z_{i-1})}  \right)
e^{-\gamma_{s_n} (T-z_{n})}.
\end{equation}
{Note also that when the trajectory is not censored , we assume that $T=Z_n$ is the entry time into the absorbing state  and the last factor in (\ref{markov2}), i.e. $e^{-\gamma_{s_n} (T-z_{n})}$,  is equal to one.}

In the semi-Markov CTMSM the transition intensity functions also depend on the time spent in the current state, that is $$ q_{rs}(t,\mathcal{F}_t)=\lim_{\delta t\rightarrow 0} \frac{  P\{Y(t+\delta t)=s|X(t)=r,T^*=t-u\}}{\delta t} $$ where $T^*$ denotes the entry time in the last state assumed before time $t$. Setting
\begin{equation*}
P\{Y(t+\delta t)= s|Y(t) = r,T^* =t-u\} =\left\{
\begin{array}{l l } 
q_{rs}(u) \delta t +o(\delta t) & s\neq r\\ 1 - \sum_{l\neq r} q_{rl
}(u) \delta t + o(\delta t) & s=r
\end{array} \right. 
\end{equation*}
we describe the whole process $Y(t)$. In fact, let $F_r(u)$ be the distribution with hazard function $\sum_{l\neq r}q_{rl}(u).$ Consider $ p_{rs}=\int_0^\infty q_{rs}(u) (1-F_r(u)) du$ and $F_{rs}(u)=\frac{1}{p_{rs}} \int_0^u q_{rs}(v) (1-F_r(v)) dv,$ for $s\neq r$. Then, $Y(t)$ is the result of the state sequence generated by the Markov chain with transition probabilities $p_{rs}$ and sojourn times depending on the departure and arrival states generated independently with distributions $F_{rs}$. To specify the functions $q_{rs} (u)$ we can also proceed directly by fixing the transition probabilities $p_{rs}$ and the conditional sojourn distributions $F_{rs}$. By doing so, the resulting hazard functions turn out to be $q_{rs}(u)={p_{rs}  F'_{rs}(u)}/{(1-F_r(u))}$ where $F_r(u)=\sum_{l\neq r} p_{rl} F_{rl}(u)$.  Several parametric can be proposed for
$q_{rs}(u)$ or $F_{rs}(u)$.  Assuming  for example cause-specific hazards proportional to those of a distribution on $(0,\infty)$ with parameters depending only on the initial state, i.e. $q_{rs}(u)=p_{rs} q(u; \theta_r) $, the transition probabilities are $p_{rs}$,
and the density of $y$ can be generally written as  
\begin{equation}
\label{smd}
p_{SM}(y)=p_{SM}(s,z)=\left(
 \prod_{i=1}^n p_{s_{i-1} s_i} q(z_{i}-z_{i-1};\theta_{s_{i-1}}) 
e^{-\int_0^{z_{i}-z_{i-1}} q(u;\theta_{s_{i-1}},) du} \right)
\times  e^{-\int_0^{T-z_{n}} q(u;\theta_{s_{n}}) du } .
\end{equation}
 % \right) \bar{F}(T-z_n;\alpha_{s_n},\gamma_{s_n})$$

The time-inhomogeneous CTMSM is obtained by considering
\begin{equation*}
P\{Y(t+\delta t)=s|Y(t)=r,\mathcal{F}_t\}=\left\{
\begin{array}{l l }
\gamma_{rs}(t) \delta t +o(\delta t) & s\neq r\\ 1+\gamma_{rr}(t)\delta
t+o(\delta t) & s=r
\end{array} \right.
\end{equation*}
In this case $Z_i$ depends on the last visited state $S_{i-1}$ and its  entry time $Z_{i-1}$, the conditional density of $Z_i|S_{i-1}, Z_{i-1}$ is
\begin{equation}
\label{fzim}
f(z_i|z_{i-1},s_{i-1})=\gamma_{s_{i-1}}(z_i) e^{-\int_{z_{i-1}}^{z_i} \gamma_{s_{i-1}}(t) dt}
\end{equation}
while $S_i$ depends on $Z_i$ and $S_{i-1}$ and 
$$P(S_i=s_i|Z_i=z_i,S_{i-1}=s_{i-1})=p(s_i|z_i,s_{i-1})=\frac{\gamma_{s_{i-1},s_i}(z_i)}{\gamma_{s_{i-1}}(z_i)}.$$
The density of $y$ can be generally written as
\begin{equation}
\label{nhmd}
p_{IM}(y)=p_{IM}(s,z)=\left( \prod_{i=1}^n p_{s_{i-1} s_i}(z_i) 
\gamma_{s_{i-1}}(z_i) e^{-\int_{z_{i-1}}^{z_i} \gamma_{s_{i-1}}(t) dt}
 \right)
 e^{-\int_{z_{n}}^{T} \gamma_{s_{n}}(t)dt }.
\end{equation}
{Finally observe that, as for the homogeneous Markov case, the last factors in the densities (\ref{smd}) and (\ref{nhmd}) are equal to one when the trajectories conclude with the entry into the absorbing state}

\section{Simulating discretely observed multi-state models}

In this Section we propose a general algorithm for simulating both semi-Markov and time inhomogeneous processes  conditionally on  the observed states $x=(x_0, x_1,\ldots, x_m, x_{m+1} )$ at times $(0,t_1,\ldots, t_{m}, T)$.  We first review the uniformization algorithm for simulating Markov processes conditional on the end-points and than we describe  how to embed the uniformization step in a Metropolis-Hastings algorithm.

\subsection{The Uniformization algorithm}

%An additional formulation for  Markov CTMSM  is based on the uniformization principle.
Let $Y(\cdot)$ be a Markov process with rate matrix $G$. 
 Consider
$\mu=\max_{r} \gamma_r$ and the transition probability matrix $$ R=I+\frac{1}{\mu} G.$$
The process $Y^*(\cdot)$ where the states follow a Markov chain  with transition matrix $R$ and the jump times a Poisson process with rate $\mu$ is equivalent to $Y(\cdot)$, see for example Ross.\cite{ross2014introduction} Note that
$Y^*$ admits virtual state changes in which a jump occurs but the state does not change. Then, to simulate the  sequence of states and jump times of $Y(\cdot)$ we can alternatively simulate $Y^{*}(\cdot)$ by drawing the Poisson process with rate $\mu$, the Markov chain sequence with transition matrix $R$ and discard the virtual jumps and the associated virtual transitions to obtain the state and jump sequence of $Y(\cdot)$.

To simulate  $Y(t)$  for $t\in [u,v]$ conditional on the end points $Y(u)=r$
and $Y(v)=s$  
Hobolth and Stone\cite{hobolth09} adopted the  uniformization  technique.  The simulation algorithm can be outlined as follows.
The first step is the simulation of the number $N$ of change points including the virtual ones whose distribution is 
$$P(N=n|Y(u)\!=\!r,\! Y(v)\!=\!s)=e^{-\mu (v-u)}\frac{(\mu (v-u))^n}{n!} R^n_{rs}/{p_{rs}(v-u)}.$$
Then we have to draw the times of the state changes $z_1^*,\ldots, z_N^*$.  Since these are realizations from a Poisson process with constant rate,   when we condition on $N=n $ they are independent and identically distributed Uniform random variable on  $[u, v ]$. Next we simulate the state sequence by a Markov chain with transition matrix $R$ conditional on the beginning state $r$ and ending state $s$, that is we simulate $Y^*(z_i^*)$, $i=1,\ldots, n-1$ from the discrete distributions 
$$ P(Y^*({z_i^*})=s_i |Y^*(z_{i-1}^*)=s_{i-1}, Y^*(t_n)=Y^*(v)=s)=\frac{R_{s_{i-1} s_i} (R^{n-i})_{s_i s}}{(R^{n-i+1})_{s_{i-1} s}}$$
where $s_o=r$.
Finally we have to discard all the virtual changes. This algorithm was
used also by Fearnhead and Sherlock\cite{fearnhead2006exact} and Pfeuffer et al.\cite{pfeuffer2018}

Note also that when  $v$ is exactly the absorbing time  we need to modify the algorithm taking account of the fact that we know that the process enter into the absorbing state at time $v$ and that  we do not know the state occupied an instant before $v$. Specifically we first need to simulate the last visited state $s$ before the absorbing time with probability
$$P_{rs}(v-u) G_{sS} \quad s=1,\ldots {S-1}$$
where $S$ denotes the absorbing state. After that we can proceed with the uniformization algorithm between $u$ and $v$ with states $r$ and $s$ where now $s$ denotes the state just simulated.

\subsection{Metropolis-Hastings for multi-state models}
To simulate the trajectories of multi-state semi Markov and inhomogeneous Markov processes conditionally on the observed states $x_0,\ldots, x_{m+1}$ at times $t_0,\ldots, T$ we   
propose a Metropolis-Hastings algorithm with proposal distribution given by the conditional Markov process. 
Let $y'=y'(t)$ for $ t\in[0,T] $ be a trajectory proposed via the conditional uniformization algorithm  previously described. This trajectory can be obtained by iterating the simulation of a Markov CTMSM with rate matrix $G$ on $[t_{i-1},t_i]$ conditioned on  $Y(t_{i-1})=x_{i-1}$ and $Y{(t_i)}=x_{i}$ for $i=1,\ldots, m+1$.  Let $(z',s')$ be the corresponding jump times and state sequence.   The proposal density of $y'$ is then $p_M(y'|x)=p_{M}(z',s')/p_M(x)$ where $p_M(z',s')$ is given by (\ref{markov}) while $p_M(x)$, which  could be obtained  multiplying the transition probabilities  $p_{x_{i-1} x_i}(t_i-t_{i-1})$, will cancel out in the acceptance ratio of the Metropolis Hastings algorithm.

Let us consider the simulation of semi-Markov trajectories. The conditional density of $y'$  under the semi-Markov model can be generally written as 
$p_{SM}(y'|x)=p_{SM}(z',s')/p_{SM}(x)$. Now $p_{SM}(x)$ cannot be neither numerically evaluated but as $p_M(x)$, it cancels out in  the acceptance ratio.  In fact let $y=y(t)$ be the current trajectory, then we accept $y'=(z',s')$ as a new 
value of the chain with probability $R=\min\left\{1,A\right\}$ where
$$A=\frac{p_{SM}(y'|x)}{p_{SM}(y|x)} \frac{p_M(y|y', x)}{p_M(y'|y ,\,x)}=
    \frac{p_{SM}(z',s')}{ p_{SM} (z, s) }  \frac{p_M(z,s)}{p_M(z',s')}.$$

%\begin{equation}
%\label{weibull}
%q_{rs}(u)=p_{rs} q(u;\alpha_r,\gamma_r)=...
%\end{equation}

Suppose, for example, that we need to simulate a conditional semi-Markov process with transition  probabilities $p_{rs}$ and Weibull sojourn times into the $r-$th state with  density $f(u;\alpha_r,\gamma_r)=  \alpha_r \gamma_r u^{\alpha_r-1} e^{-\gamma_r u^{\alpha_r}}$  for $r=1,\ldots, S$. The rates of the Markov proposal may be assumed to be either equal to the rate parameters of the Weibull semi-Markov model or equal to the inverse mean of the Weibull semi-Markov process. Anyway there are no substantial differences in the approaches. In both the cases as $\alpha_r\to1$ for each $r$, the proposal converges to the target distribution and $A\to 1$. Assuming for simplicity the first setting, we propose a trajectory from a Markov process conditional on $x$ with the same transition probabilities $p_{rs}$ and rate parameters $\gamma_r$, which is accepted with ratio given by 
\begin{eqnarray*}
A=\frac{p_{SM}(z',s')}{p_M(z',s')  }  \frac{p_M(z,s)}{p_{SM} (z, s)}&=&
\prod_{i=1}^{n'} \frac
{  \alpha_{s'_{i-1}} 
%\gamma_{s_{i-1}}  
(z'_i-z'_{i-1})^{\alpha_{s'_{i-1}}-1} e^{-\gamma_{s'_{i-1}} (z'_i-z'_{i-1})^{\alpha_{s'_{i-1}}}}} 
{ 
%\gamma_{s_{i-1}} 
e^{-\gamma_{s'_{i-1}}(z'_i-z'_{i-1})} }
\times 
\frac{e^{-\gamma_{s'_{n'}} (T-z_{n'})^{\alpha_{s_{n'}}}}}{e^{-\gamma_{s_{n'}} (T-z_{n'})}}
\\
&\times&
\prod_{i=1}^n \frac{ 
%\gamma_{s_{i-1}} 
e^{-\gamma_{s_{i-1}}(z_i-z_{i-1})} }
{  \alpha_{s_{i-1}} 
%\gamma_{s_{i-1}}  
(z_i-z_{i-1})^{\alpha_{s_{i-1}}-1} e^{-\gamma_{s_{i-1}} (z_i-z_{i-1})^{\alpha_{s_{i-1}}}}} 
\times 
\frac{e^{-\gamma_{s_n} (T-z_{n})}}
{e^{-\gamma_{s_n} (T-z_{n})^{\alpha_{s_n}}}}.
\end{eqnarray*}
{ Note also that the ratios between the survivor functions of the Weibull and the Exponential random variables in the previous formula formula appear only in the case of uncensored trajectories.}

To simulate the conditional trajectories in the  inhomogeneous case we {suggest} to proceed in the same way as in the semi-Markov case by proposing from a Markov model with rates taken by the corresponding matrix of the inhomogeneous model calculated in a point $t^*$ internal to the interval $[0,T]$. %An alternative approach may be to propose from a piecewise homogeneous Markov model, with rate matrix stepwise dependent on time: however, this solution increases the computational complexity of the model. By homogeneous proposal the convergence of the Markov Chains is fast for each of the parameters and the time dependence of the rate matrix is captured without any problem. 
For example suppose that the rate functions of the inhomogeneous model are given by
\begin{equation}
\label{Gomperz}
\gamma_{rs}(t)=p_{rs} e^{\beta_{0r}+\beta_{1r} t}, \quad r,s=1,\ldots S.
\end{equation}
In this case the conditional density (\ref{fzim}) becomes
$$f(z_i|z_{i-1},s_i)=e^{\beta_{0 s_{ i-1} }+\beta_{1 s_{i-1} } z_i }
 e^{-\frac{e^{\beta_{0 s_{i-1}}}}{\beta_{1 s_{i-1}}}
\left[e^{\beta_{1 s_{i-1} } z_i} -e^{\beta_{1 s_{i-1} } z_{i-1}}\right] }
 $$
 and the acceptance ratio is given by
 \begin{eqnarray*}
A=\frac{p_{IM}(z',s')}{p_M(z',s')  }  \frac{p_M(z,s)}{p_{IM} (z, s)}&=&
\prod_{i=1}^{n'} \frac
{ e^{\beta_{0 s'_{i-1} }+\beta_{1 s'_{i-1} } z'_i }
 e^{-\frac{e^{\beta_{0 s'_{i-1}}}}{\beta_{1 s'_{i-1}}}
\left[e^{\beta_{1 s'_{i-1} } z'_i} -e^{\beta_{1 s'_{i-1} } z'_{i-1}}\right] }} 
{ 
%\gamma_{s_{i-1}} 
 \gamma_{s'_{i-1}} e^{-\gamma_{s'_{i-1}}(z'_i-z'_{i-1})} }
\times
\frac{e^{-\frac{e^{\beta_{0 s'_{i-1}}}}{\beta_{1 s'_{i-1}}}
\left[e^{\beta_{1 s'_{n'} } T} -e^{\beta_{1 s'_{n'} } z'_{n'}}\right] }}
{e^{-\gamma_{s_{n'}} (T-z_{n'})}}\\
&\times&
\prod_{i=1}^n \frac{ 
%\gamma_{s_{i-1}} 
\gamma_{s_{i-1}} e^{-\gamma_{s_{i-1}}(z_i-z_{i-1})} }
{  e^{\beta_{0 s_{i-1} }+\beta_{1 s_{i-1} } z_i }
 e^{-\frac{e^{\beta_{0 s_{i-1}}}}{\beta_{1 s_{i-1}}}
\left[e^{\beta_{1 s_{i-1} } z_i} -e^{\beta_{1 s_{i-1} } z_{i-1}}\right] }} 
\times 
\frac{e^{-\gamma_{s_n} (T-z_{n})}}
{
e^{-\frac{e^{\beta_{0 s_{n}}}}{\beta_{1 s_{n}}}
\left[e^{\beta_{1 s_{n} } T} -e^{\beta_{1 s_{n} } z_{n}}\right] }
}
\end{eqnarray*}
where $\gamma_r=\exp(\beta_{0r}+\beta_{1r} t_0).$

Note that when the rate functions $\gamma_{rs}(t)$ are strongly time dependent we may alternatively propose from a piecewise homogeneous Markov model with change points chosen accordingly to the observations intervals $[t_{i-1}, t_{i}]$ for $i=1,\ldots, m+1$ and rate parameters $\exp(\beta_{0r}+\beta_{1r} t^*_{i})$ where $t^*_i$ is an internal  point of the interval $[t_{i-1},t_{i}]$. { However, this solution increases the computational complexity of the model.}

%%%% IPOTESI 1: UTILIZZARE UN  ALGORITMO CHE SCHEMATIZZA  SOLO IL METROPOLIS HASTINGS PER TRAIETTORIA SINGOLA
{ Let $\theta$ be the parameter vector of either the semi-Markov or the inhomogeneous Markov multi-state model. For each observed individual, let $(s,w)^{(j-1)}$ be the last accepted trajectory at the iteration $j$. The Metropolis-Hastings for multi-state paths simulation works as follows:
\begin{itemize}
\item set the rate matrix $G$ as function of $\theta^{(j-1)}$;
\item draw $(s',z')$ from $\text{Uniformization}((s,w)^{(j-1)},G)$ ;
\item draw $\omega$ from $U(0,1)$;
\item set $(s,z)^{(j)}=(s',z')$ if $\omega<A$; else $(s,z)^{(j)}=(s,z)^{(j-1)}$.
\end{itemize}
}
%%%% IPOTESI 2: UTILIZZARE UN  ALGORITMO CHE SCHEMATIZZA  SOLO IL METROPOLIS HASTINGS PER N TRAIETTORIE 
%\begin{itemize}
%\item set the rate matrix $G$;
%\item for $i=1,\dots,N$
%\begin{itemize}
%\item set the rate matrix $G$;
%\item draw $(s',z')_i$ from $\text{Uniformization}((s,w)_i^{(j-1)},\theta^{(j-1)})$ ;
%\item draw $\omega$ from $U(0,1)$;
%\item set $(s,z)_i^{(j)}=(s',z')_i$ if $\omega<A$; else $(s,z)_i^{(j)}=(s,z)_i^{(j-1)}$.
%\end{itemize}
%\end{itemize}

\subsection{The full MCMC algorithm for panel data}

Now we suppose to have $n$ discretely observed  realizations from a multi-state model.  Let $x_i=(x_{i\,0},$ $x_{i\,1},\ldots,x_{i,m},x_{i\,{m_i+1}})$ be the observed states at the times 0=$t_{0}<t_{i,1}<\cdots<t_{i,m}<t_{i,m_i+1}=T_i$ for the {\em i-th} unit. 
We assume that the observation times can be irregularly spaced and may also be unequal for all the sample units.
Moreover we assume that the first state  $x_{i\,0}$ is fixed for $i=1,\ldots, n$. Finally let $p(y|\theta)$ be the generic density for a complete observed trajectory $y$ where $\theta$ represent the unknown parameter vector.

To make Bayesian inference for the  parameter $\theta$
under the  semi-Markov and inhomogeneous Markov models we may  perform a Metropolis within Gibbs algorithm  where at each step  of the chain we update the parameter $\theta$ on the base of the conditional distribution
$$p(\theta|y_1,\ldots, y_n) \propto  \prod_{i=1}^n  p(y_i|\theta) p(\theta)$$
given the whole trajectories $y=(y_1,\ldots,y_n)$.  Note that, the simulation of  the components of $\theta$ will depend 
on  the parametric family of the density $p(y|\theta)$ and the prior distribution $p(\theta)$ and  will be performed 
via  standard  Metropolis-Hasting updating  or directly by Gibbs step if it is possible. Then we update 
the trajectories $y_i$ for $i=1,\ldots, n$  by  the Metropolis-Hastings  step described in the previous sub-section.

Note that, for the updating of the  paths $y_i$ for $i=1,\ldots, n$, we suggest to use as proposal distribution a Markov process with rate matrix dependent on the last value of  $\theta$  generated from the Markov chain.. %%% Secondo me dobbiamo limitarci a suggerire di usare un modello che dipenda da theta, perche' a livello teorico, seppur lenta la convergenza dovrebbe esserci anche con dei parametri fissi per la proposal.
For example, in the Weibull semi-Markov model with parameters  $\theta=\left(P,\alpha,\gamma \right)$
where $P$ is the matrix with the transition probabilities while  $\alpha=(\alpha_1, \ldots, \alpha_S)$ and $\gamma=(\gamma_1,\ldots, \gamma_S)$ are the Weibull parameters, 
 we may use as proposal distribution a Markov process with  rate  matrix $G$ having on the main diagonal the  values $\gamma_{rr}=-\gamma_r$ for $r=1,\ldots, S$ where $\gamma_r$ are the current values for the rate parameters of the Weibull holding times
 and off-diagonal elements $\gamma_{rs}=\gamma_r p_{rs}$ where $p_{rs}$ are the current values for the transition probabilities.

 Similarly,  in the inhomogeneous case with $\theta=\left(P,\beta_{0},\beta_{1} \right)$, where $\beta_0=(\beta_{01}\,\ldots,\beta_{0S})$ and $\beta_1=(\beta_{11},\ldots, \beta_{1S})$,  we may take the last simulated values of $P, \beta_0$ and $\beta_1$ and propose from a Markov process with rate matrix $G$ of elements $\gamma_{rr}=-\exp{\left(\beta_{0r}+\beta_{1r}t_{0_i}\right)}$ and $\gamma_{rs}=p_{rs}\exp{\left (\beta_{0r}+\beta_{1r}t_{i}^* \right )}$, where  $t_{i}^*\in \left [0,T_i \right ]$.

Moreover, if $\alpha_r=1$ for each $r\in S$ in the semi-Markov case or $\beta_{1r}=0$ for each $r\in S$ in the time-inhomogeneous Markov case, our algorithm allows to simulate from the true conditional distribution of the process trajectories. Hence, the closer are the models to the nested Markov case, the closer are the proposals to the true conditional distributions. However, when the  departure from the Markov model  is larger, the conditioning to the endpoints in each sub-interval helps in generating trajectories similar to the latent ones.

%%%% IPOTESI 3: UTILIZZARE UN UNICO ALGORITMO CHE SCHEMATIZZA L'INTERO QUADRO MCMC 
%The general MCMC sampling scheme may be summarized as follows:
%\begin{itemize}
%\item \textit{Simulating paths}
%\begin{itemize}
%\item $(s',z')\sim\text{Uniformization}(\theta^{(j-1)})$ ; $\omega\sim U(0,1)$;
%\item accept if $\omega<A$;
%\end{itemize}
%\item \textit{Updating model parameters}
%\begin{itemize}
%\item $\theta^{(j)}\sim p(\theta|y)$.
%\end{itemize}
%\end{itemize}

%Forse dovremmo dire che quando $\alpha=1$ o $\beta_1=0$ stiamo simulando dalla vera condizionata del processo e che quando stiamo vicini a questi valori la nostra proposta è molto simile alla vera condizionata e che poi quando ci allontaniamo comunque il condizionamento agli stati endpoint di ogni sottointervallo aiuta a generare traiettorie simili a quelle latenti

\section{Applications}

%In this section we present results with both simulated and real data. In particular we consider two data sets for multi state models, also comparing results with the ABC approach by \cite{tancredi2019biom}.

\subsection{Breast cancer data}

We first consider a data set comprising 37 women with breast cancer treated
for spinal metastases; see De Stavola \cite{de1988testing}, Davison \cite{davison2003statistical} and Tancredi \cite{tancredi2019biom} for
previous analyses of this data set. The ambulatory status of the
women, defined as ability to walk unaided or not, was recorded when
the treatment began and then 3, 6, 12, 24, and 60 months after
treatment. The three states are: able to walk unaided (1) unable to
walk unaided (2); and dead (3). 
%Thus a sequence $x_0=1,x_1=1,x_2=1,x_3=3$ means that the patient was able to walk
%unaided each time she was seen during the first year follow-up but
%was dead before the end of the second year.  
Note that several sequences are censored. For example, for a  patient we have only 
$x_0=1,x_1=1,x_2=1,x_3=2$ meaning  that the patient was not seen after the first year. For these patients the proposed algorithm reconstruct the unknown trajectory until the censoring time. 
{Moreover in this application, the death time is not exactly observed. For patients whose last observation is the dead state, the algorithm simulates the death time  in the last observational interval conditionally on the end-point states, with the right one being the death. }

Following the analysis conducted in Davison \cite{davison2003statistical} we illustrate  the results obtained both assuming $p_{13}=0$, i.e. the impossibility to have instantaneous transitions from state 1 to the death state, and permitting these transitions to occur, i.e. $p_{13}>0$. 
These two assumptions are considered both under a standard Markov model   and under the proposed Weibull semi-Markov model. Note also that, contrary to Davison \cite{davison2003statistical}, the starting observations at time 0 were assumed fixed.  As a prior distribution for the  parameters of the semi-Markov process we take  a standard Normal for $\log(\alpha_i)$, $i=1,2$. Moreover we consider  Gamma distributions with hyper-parameters $f=0.001$ and $g=0.001$ for the parameters $\eta_i=\gamma_i^{\alpha_i}$, $i=1,2$ and Uniform distributions for the unknown  transition probabilities. For each of the four models we drew the posterior distribution of the model parameters running the MCMC algorithm discussed in the previous Section for 50000 iterations.

Table  \ref{destavola1}  reports the posterior means, the posterior standard deviations and the 95\% posterior credibility intervals obtained under these settings.  We provide also the posterior summaries of the model parameters obtained 
by the approximate Bayesian computation (ABC) approach of Tancredi \cite{tancredi2019biom}. 
In comparing the proposed methodology with respect to the ABC solution we note that  the updating of the latent trajectories slow down the mixing of the MCMC algorithm. Anyway, due to the lack of an analytical expression for the likelihood function of the semi-Markov models, the trajectories reconstruction and the consequent slow mixing is actually unavoidable for producing MCMC  inference for this class of models. 
However,  the posterior results in this example are quite robust with respect to two computational strategies, although, ABC inference, as expected, slightly overestimates the posterior uncertainty as illustrated by the wider posterior intervals.
%both for the rate parameters of the Markov and semi- Markov models and the shape parameters  of the semi-Markov models reported in Table \ref{destavola1}. 

\begin{figure}
\centerline{
\includegraphics[width=18cm]{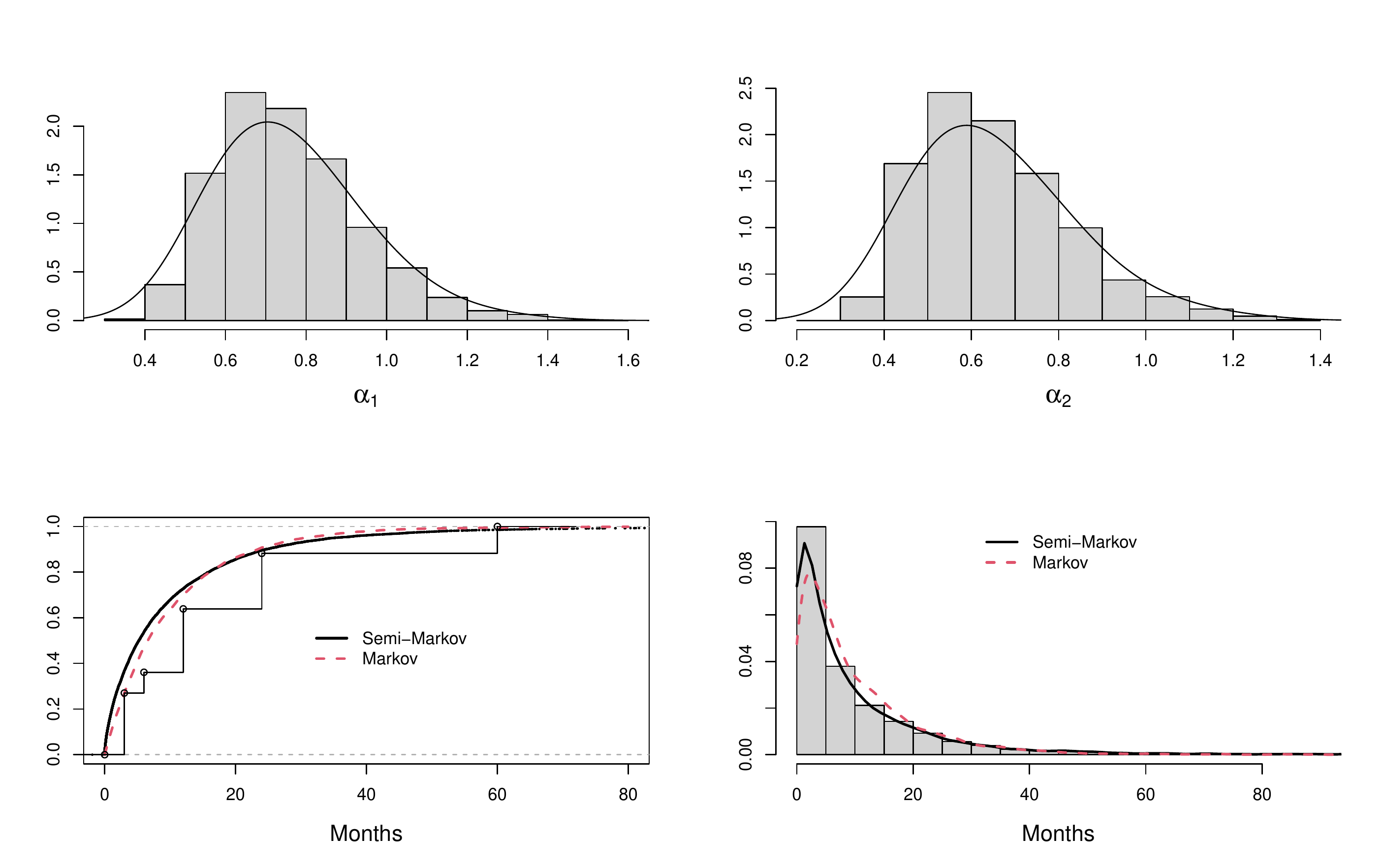}}
\caption{Breast cancer data set. Upper panels: posterior distributions for the shape parameters of the semi-Markov model with $p_{13}=0$. Lower panels: posterior predictive distributions of the death time (cumulative and density) for the semi-Markov and Markov models. \label{detavola_fig} }
\end{figure}

Finally, the upper panels of Figure \ref{detavola_fig} show the MCMC posterior distributions of  the shape parameters $\alpha_1$ and $\alpha_2$ for the semi Markov model with $p_{13}=0$ while
 the lower panels  report the cumulative posterior predictive distribution
and the corresponding predictive posterior density of  the death time for both the Markov and semi-Markov models with $p_{13}=0$.  Note that both the 95\% credibility intervals of the shape parameters contain the value $\alpha=1$ showing no evidence to reject the Markov model. 
 Anyway the semi-Markov model seems to produce slightly ticker tails for the death time predictive distribution reflecting the tendency of the posterior distributions for $\alpha_1$ and $\alpha_2$ to concentrate on values less than 1.

\subsection{Cardiac allograft vasculopathy data}

We now consider the data used by Sharples et al\cite{sharples2003diagnostic} to analyze the progression of coronary allograft vasculopathy (CAV), a disease leading to the deterioration of arterial walls which is a common cause of death after heart transplantation. The CAV data are available with the R package \texttt{msm},\cite{jackson2011multi} and provide the disease status (CAV-free (1), mild CAV (2) and moderate or severe CAV (3)) observed approximately each year after transplant for a set of 622 subjects followed up until their most recent visit if alive at the end of the observation period or until death (state (4)). Death times are exactly observed. The data also comprise apparent transitions from higher to lower states, which are in fact the results of a misclassification since the deterioration of the arterial walls is an irreversible process.  
For this application  we  fitted both semi-Markov and time-Inhomogeneous Markov models to the data obtained by progressively recoding all the reverse transitions as remaining in the higher of the two states and permitting transitions only to the adjacent states or to the death state.

%We first present the results for the Markov and semi-Markov model.   We consider the same prior model of the breast cancer data and we run our MCMC algorithm for 10000 iterations.  In Table \ref{CAV1} we report the posterior means, the posterior standard deviations and the 95\% posterior credibility intervals. Notice that the rate parameters and the transition probabilities do not differ significantly across the two models. In fact  all the credibility intervals for Weibull shape parameters include $\alpha=1$. (dire quindi che non c'è evidenza per i semi-Markov)

%The lower part of Table \ref{CAV1} shows the results for the time inohomogeneous model. Note that we observe a significant time effect for the sojourn time in the second state ,but note for the other. 

\begin{figure}
\centerline{
\includegraphics[width=18cm]{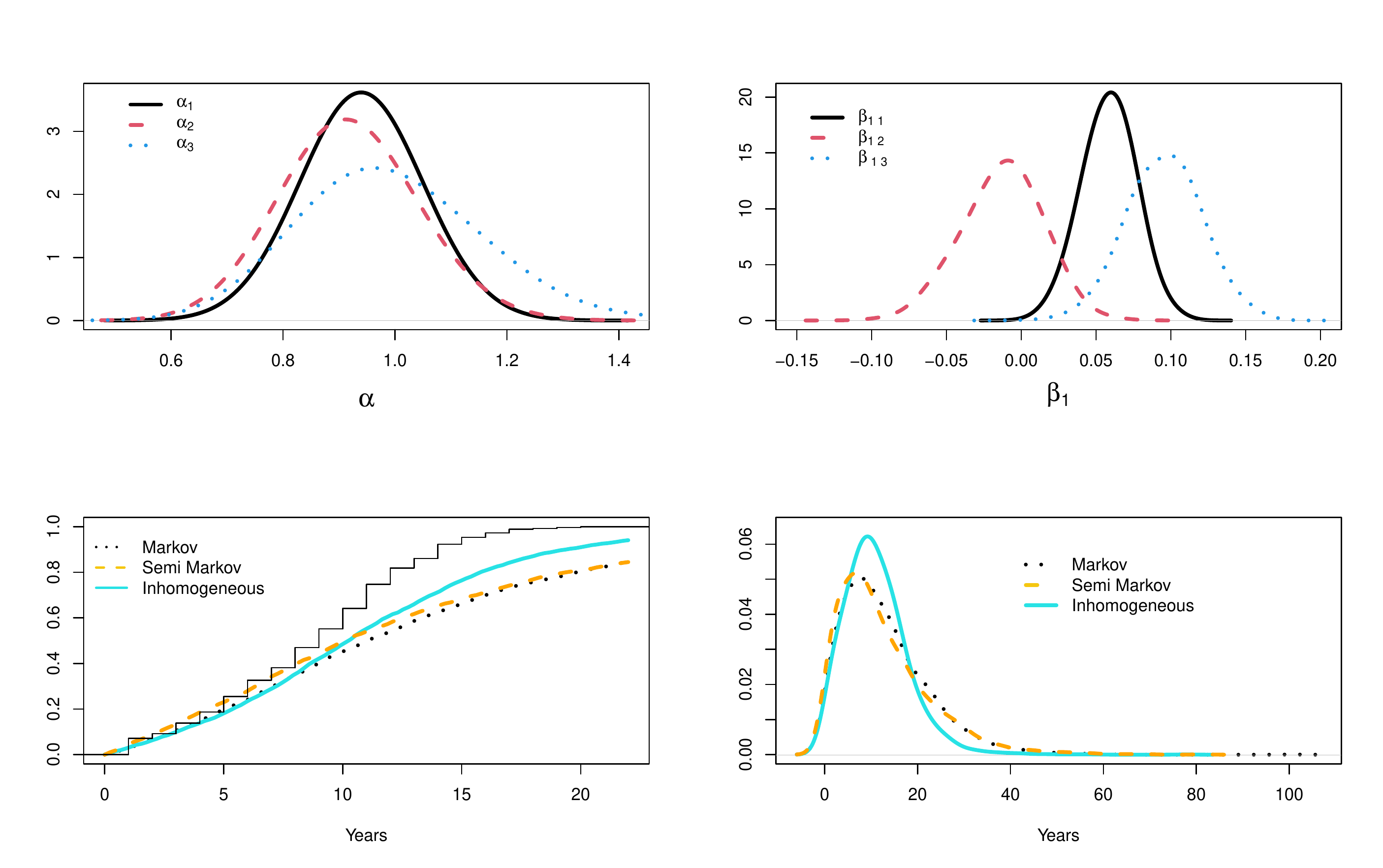}}
\caption{CAV data set. Upper panels: posterior distributions for the shape parameters of the semi-Markov model and for the time effect paramters of the inohomogeneous models. Lower panels: posterior predictive distributions of the death time (cumulative and density) for the Markov, semi-Markov and inhomogeneous models. \label{cav_fig} }
\end{figure}

For the Markov and semi-Markov models we consider the same prior setting used in the previous application, while for the inohomogeneous model we define the rate function as in (\ref{Gomperz}) with diffuse Normal prior for the time regression parameters. For the transition probability matrix we use the same prior as in the other models. We run our MCMC algorithms for 10000 iterations. In Table \ref{CAV1} we report the posterior means, the posterior standard deviations and the 95\% posterior credibility intervals. Notice that the rate parameters and the transition probabilities do not differ significantly across the Markov and semi-Markov models. In fact all the credibility intervals for Weibull shape parameters include $\alpha=1$. Also in this case there is no semi-Markov evidence. Instead, we observe a significant time effect for the sojourn times in the first and third states. In the upper panel of Figure \ref{cav_fig} we show again the MCMC posterior distributions of the shape parameters for semi-Markov model and the time dependence parameters for the inhomogeneous model. In the lower panels we report the the death time predictive distributions for the Markov, semi-Markov and inhomogenous models. Note that the inhomogenous model fits the empirical death state occupation probability better than the other models.

\begin{figure}
\centerline{
\includegraphics[width=4cm]{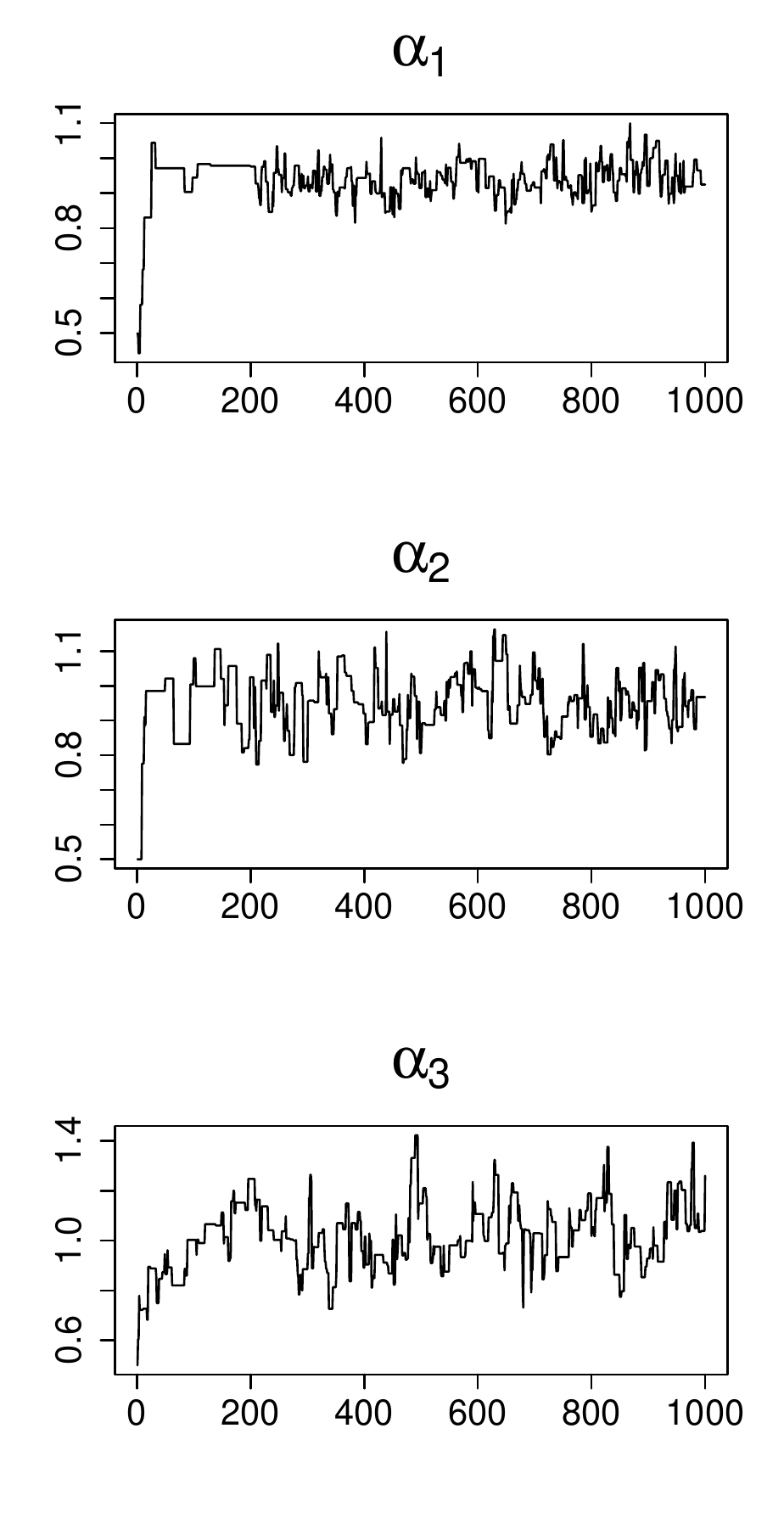}
\includegraphics[width=4cm]{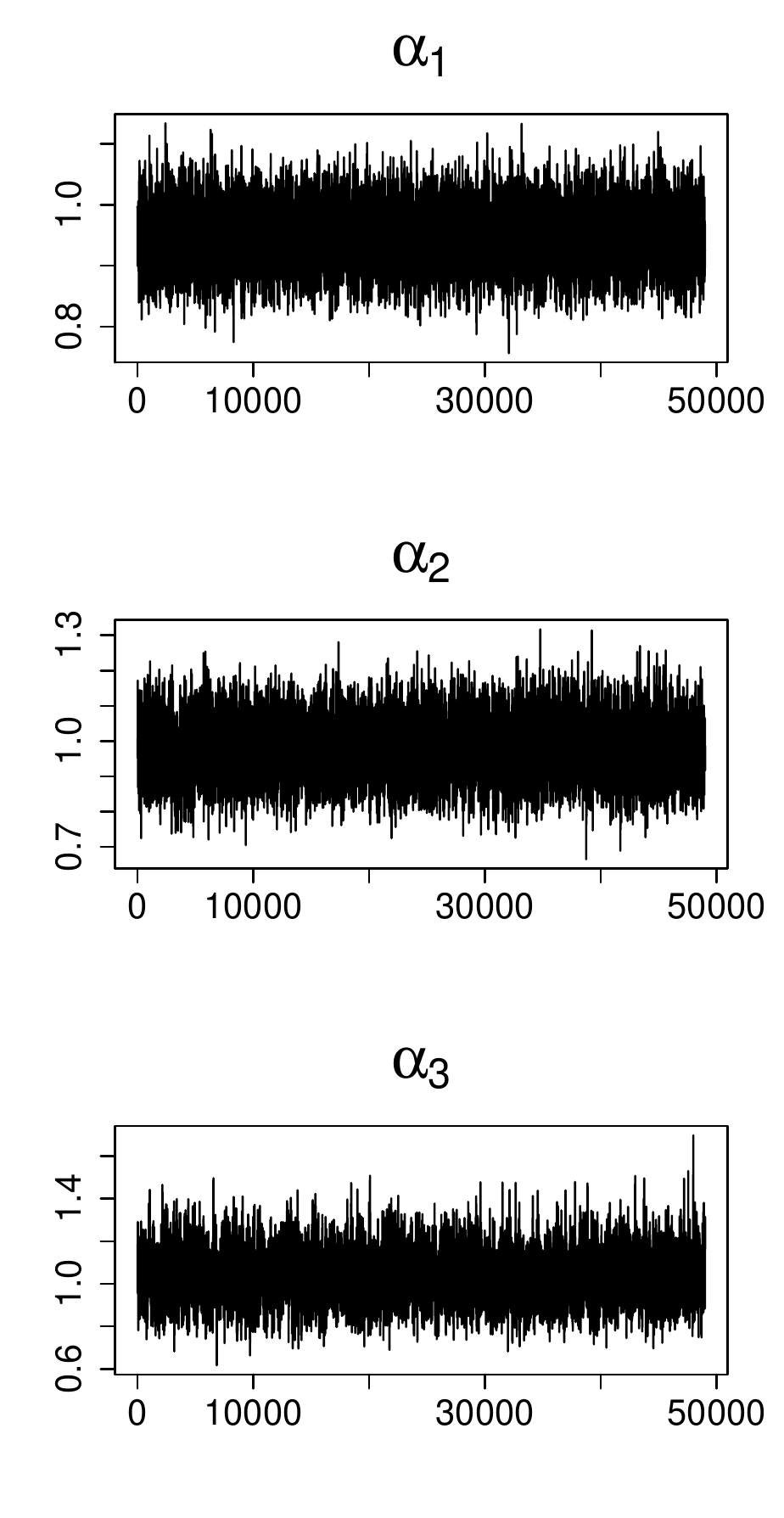}
\includegraphics[width=4cm]{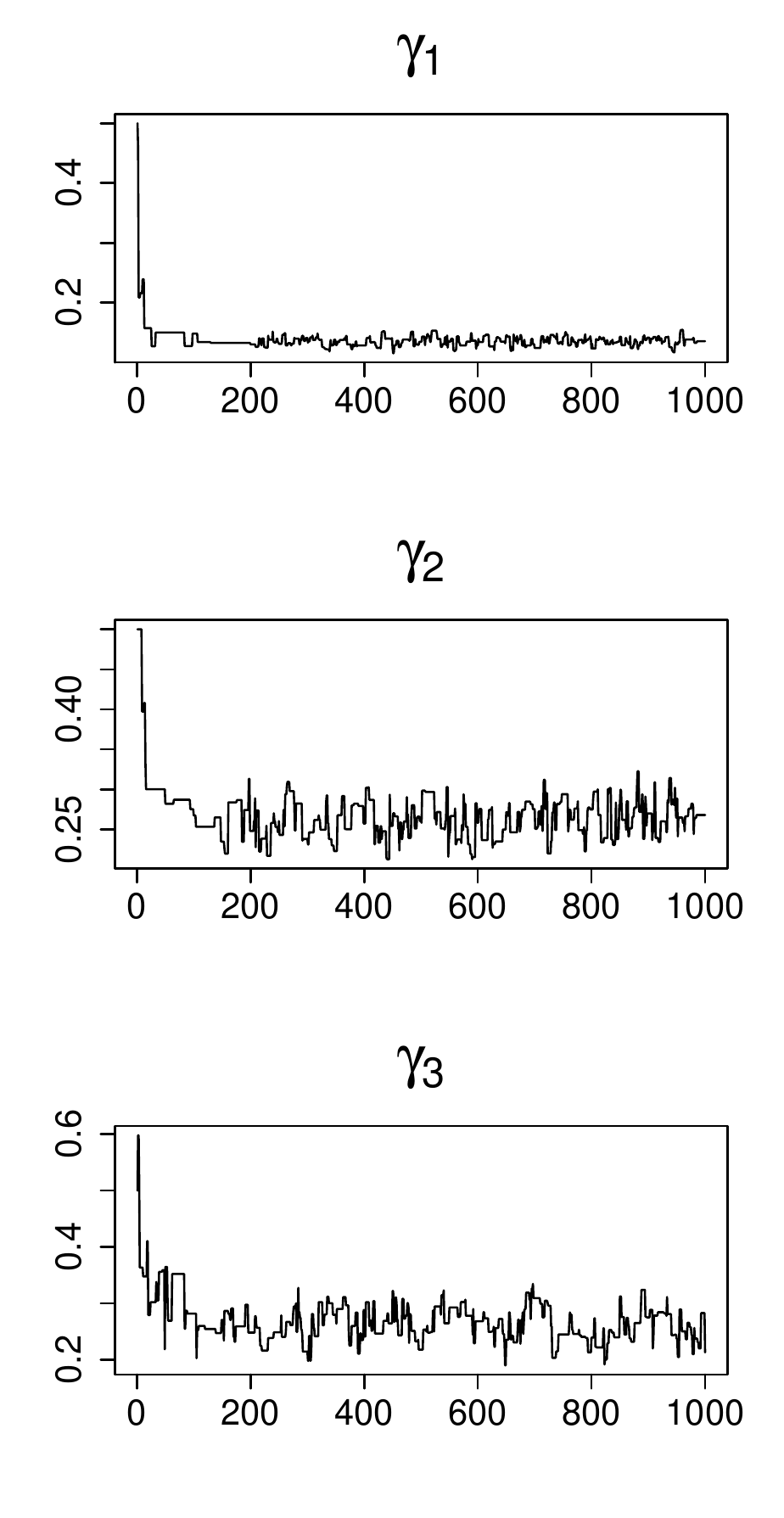}
\includegraphics[width=4cm]{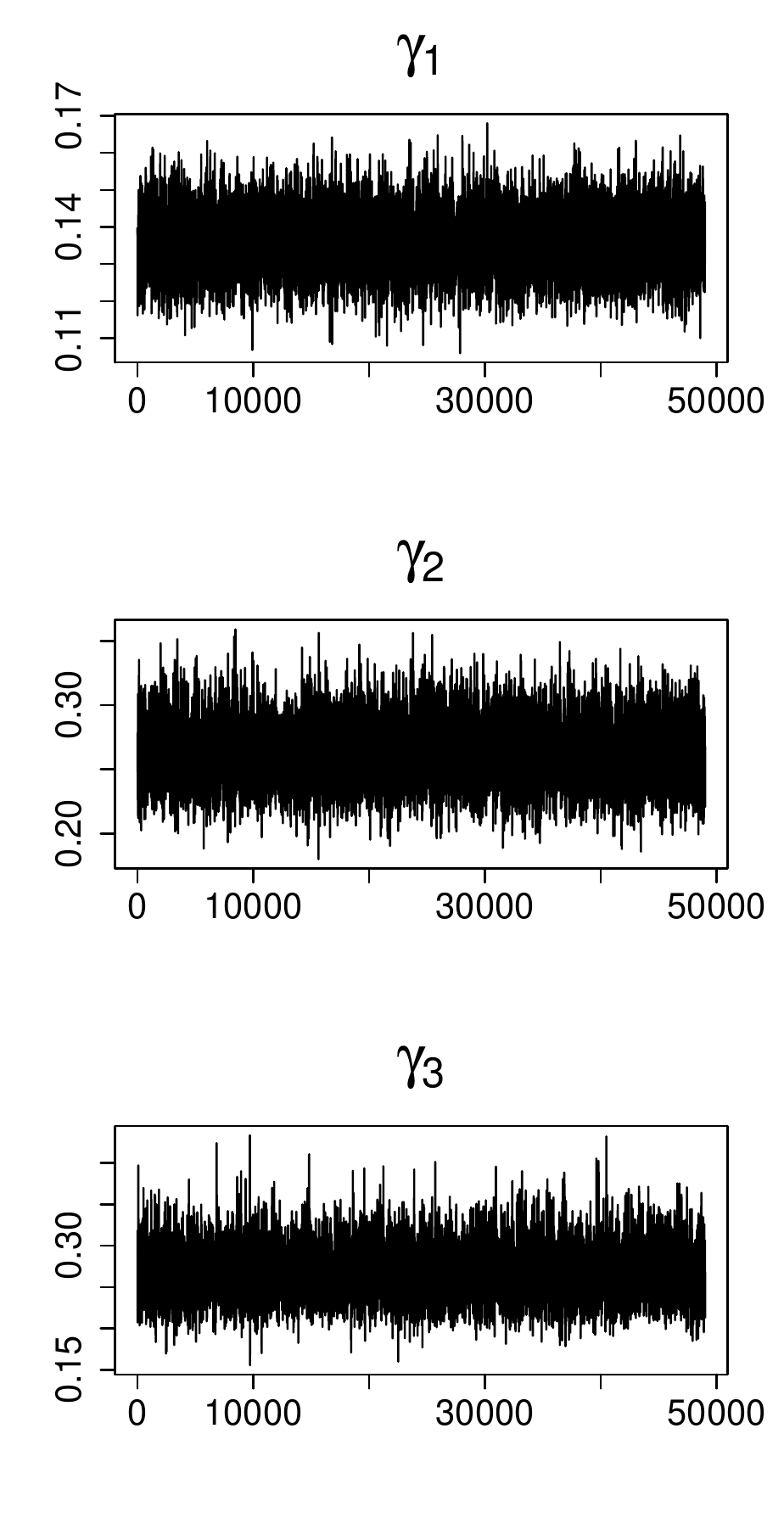}}
\centerline{
\includegraphics[width=4cm]{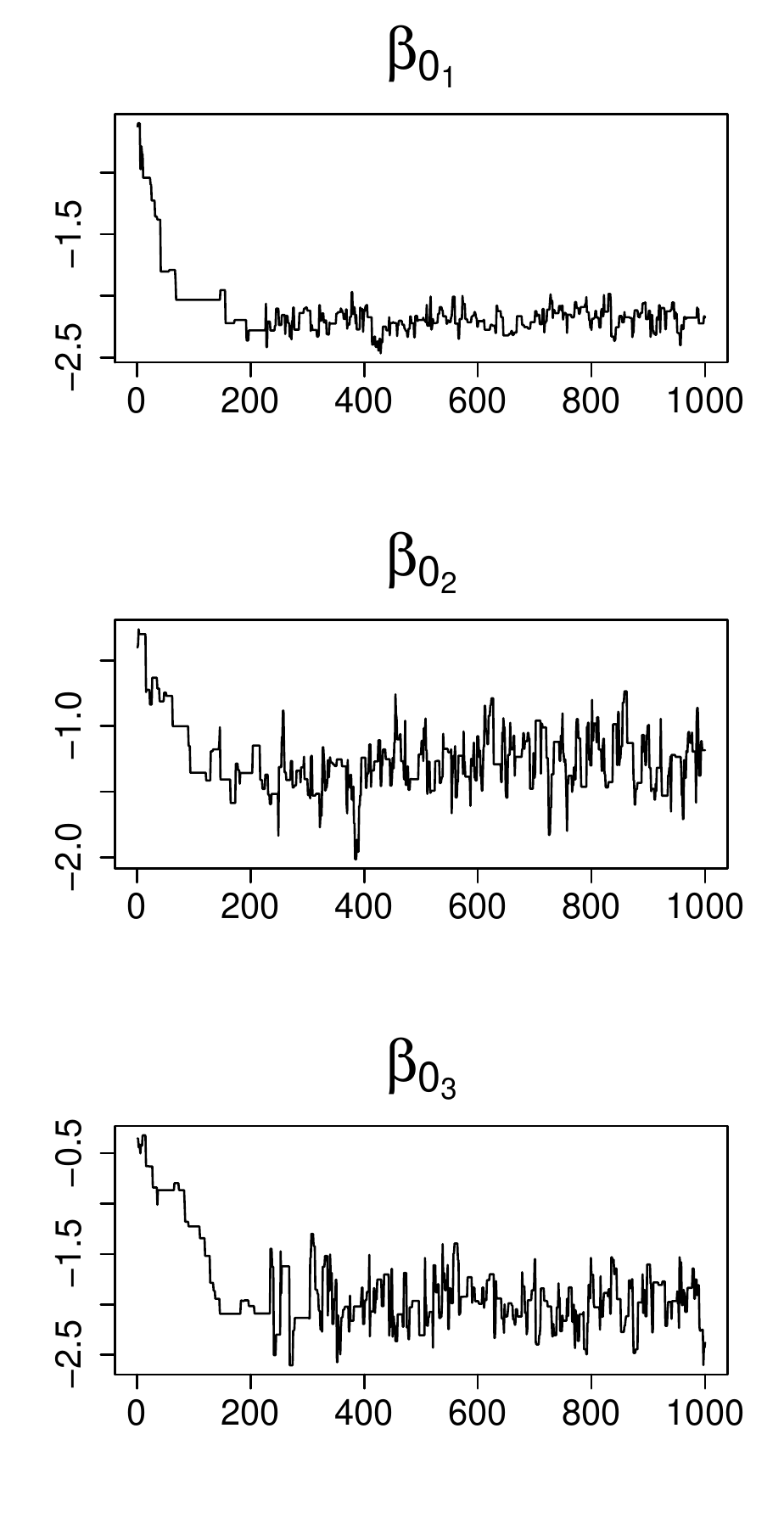}
\includegraphics[width=4cm]{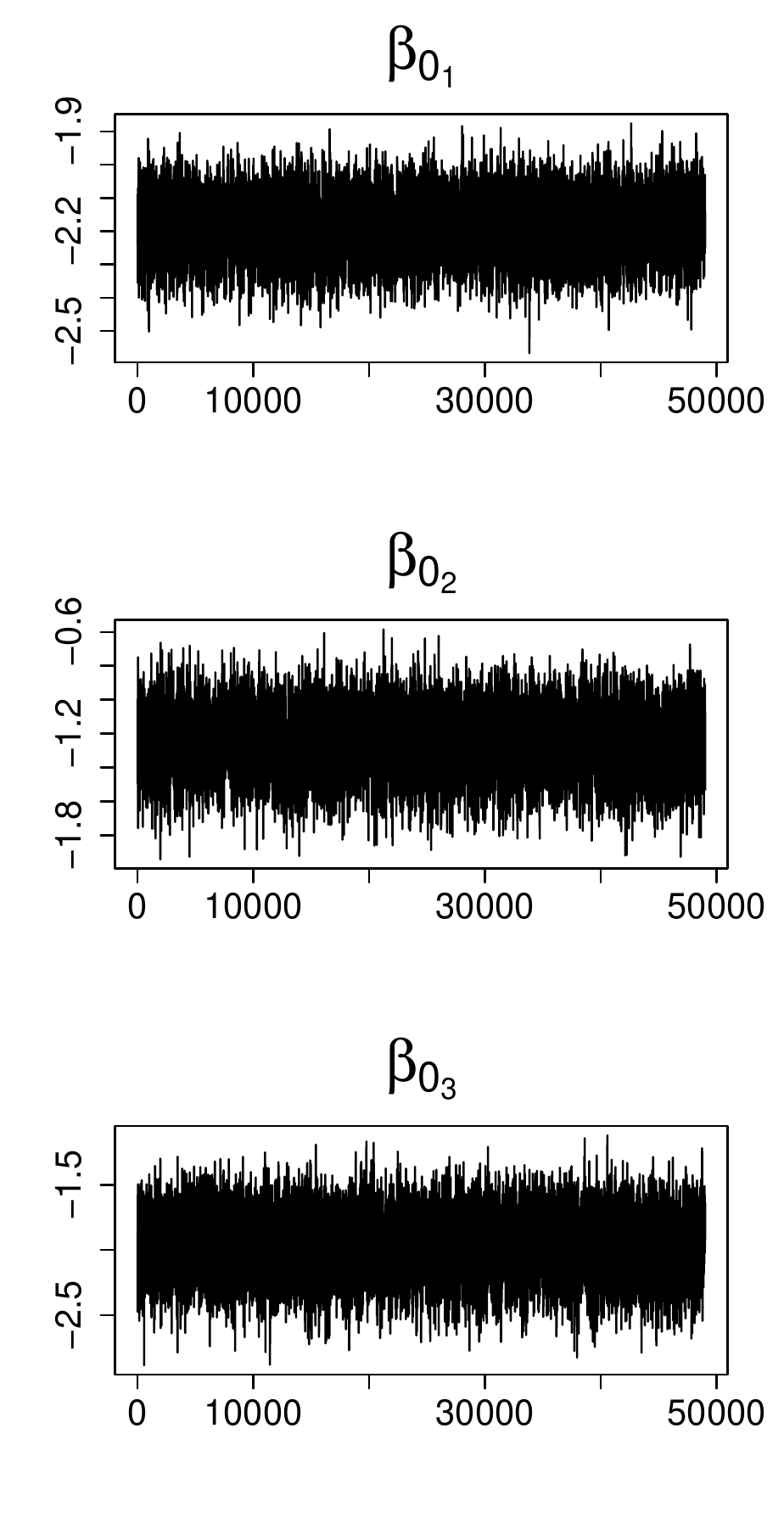}
\includegraphics[width=4cm]{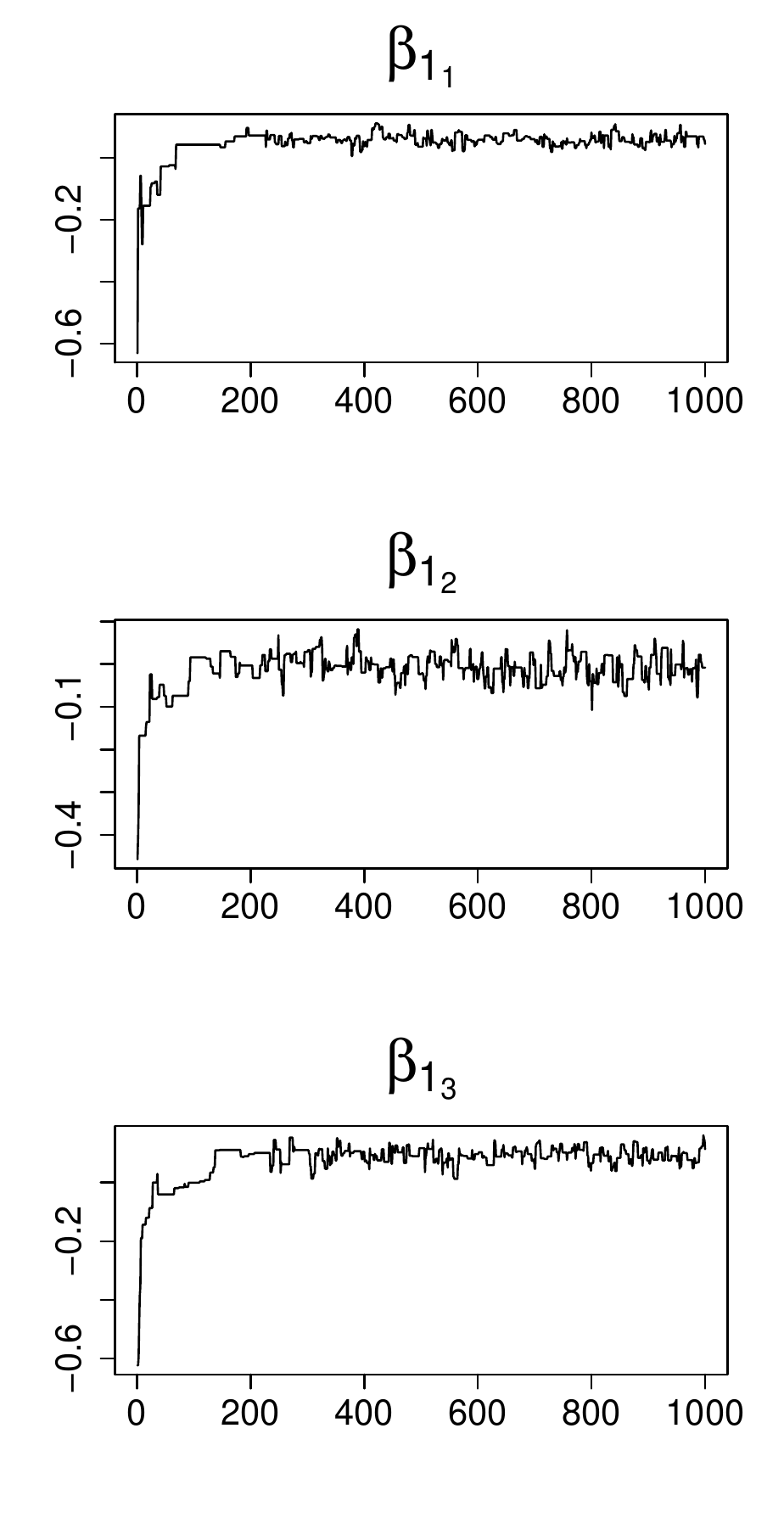}
\includegraphics[width=4cm]{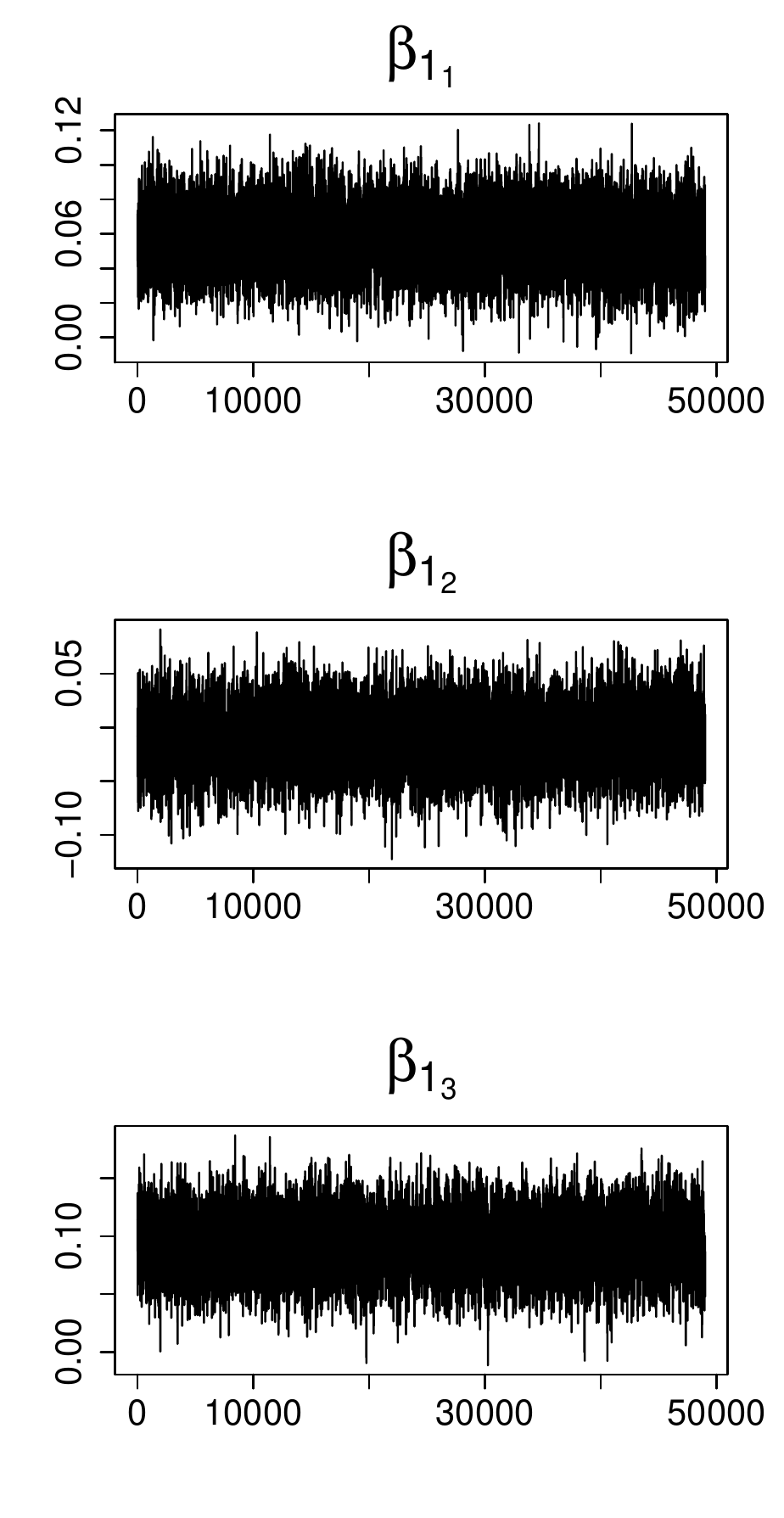}}
\caption{CAV data set. Traceplots of shape and rate parameters for the Weibull semi-Markov model and the time inhomogeneous Markov model. \label{MCMCtraces} } 
\end{figure}

Lastly, we provide the MCMC diagnostics for both the models. As can be seen from the traceplots in Figure \ref{MCMCtraces}, despite the model compleixty, the convengerce is fast. Moreover, the traces are stable over 50000 iterations.

%\newpage

\subsection{Simulation study}
To assess the proposed methodology, we applied the MCMC algorithms to simulated data sets, partially replicating the experiment conducted by Titman\cite{titman2014estimating} and Tancredi.\cite{tancredi2019biom} In particular, for the semi-Markov model, data were generated from a model with three states: healthy, ill, dead. All patients start in the healthy state and can recover from the ill state according to a Weibull model with transition intensity functions $q_{rs}(u)=\gamma_{rs} \alpha_r (u\gamma_r) ^{ \alpha_r-1}$ where $\gamma_r=\sum_{s\neq r} \gamma_{rs}$. The exact model parameters are fixed to $\theta=$ $(\gamma_{12},\alpha_1,$ $\gamma_{13}, $ $ \gamma_{21}, $ $\alpha_2, $ $ \gamma_{23})$ $=$ $(0.25,1.4,0.05,0.04,0.7,0.1)$, corresponding to a process where the hazard of the transition out from the state is increasing with time for the healthy state and decreasing for the ill state. Moreover, the transition probability towards the dead state is greater under the ill state ($p_{23}=\gamma_{23}/\gamma_2=0.71$) than with the healthy state ($p_{13}=\gamma_{13}/\gamma_1=0.167$).   Moreover  note  that the follow-up times are set equal to (0,3, 6,12,24,60) months and that we consider both the cases with the death times unknown and known.  We used the same prior distribution of the previous example.  We set the sample size at $n=50,100,500$ and 1000 and for each sample size we generated 100 data sets running the MCMC algorithm for 10000 iterations. Table \ref{tab_sim_SM} reports the empirical averages and standard deviations of the posterior means obtained for each simulated data set. Note that  increasing the sample size led the Bayesian estimators to concentrate on the true values of the parameters both when the detah time is unknown (upper table) and when it is know. We notice also that, as expected, the information introduced by assuming that the death time is exactly known provides
always a smaller mean square error with respect to the unknown death time scenario.

For the time-Inhomogeneous Markov model, data were always generated from a three state model.  Again, patients start in the healthy state (1) and can recover from the ill state accordingly to the transition intesities  $\gamma_{rs}(t)=p_{rs}\gamma_{r}(t)$  given by  (\ref{Gomperz}) .We use the same prior setting as in the CAV data example. The exact model parameters are $\theta=(\beta_{0_1},\beta_{1_1},\beta_{0_2}, \beta_{1_2}, p_{12}, p_{13},p_{21},p_{23})=(-0.69,0.2,-2.30,0.2,0.8,0.2,0.2,0.8)$ and follow-up times are equal to (0,1,2,3,4.5,6,9,12,20), with known death time. In particular, we have have assumed that the transitions rates are increasing with time. We generated 100 data sets of size $n=50,100,500$ and 1000, running the MCMC algorithm for 10000 iterations. Table \ref{tab_sim_TIM} reports the results. Also in this case we observe that as the sample size increases the Bayesian estimators concentrate on the true values of the parameters.

\section{Discussion}

In this paper we have shown how the uniformization technique for the simulation of condtioned endpoints continuous time Markov trajectories can be embedded in a Metropolis-Hastings algorithm to perform Bayesian inference for discretely observed semi-Markov and time-inhomogeneous Markov models.
In the applications we considered Weibull sojourn time distributions  for the semi-Markov models and a Gomperz link function for the time-inhomogeneous Markov model, but the proposed approach can be easily extended to any type of sojourn distributions or rate link functions. 

{ To compare from a general point of view our MCMC approach with respect to other numerical techniques for estimating discretely observed multi state models
%models like the likelihood approximations of \cite{titman2011flexible} and  \cite{machado2021penalised} or the ABC approach of \cite{tancredi2019biom} 
we notice that the MCMC formalization may be easily extended to a broader class of multi-state models. In fact  once the individual latent trajectory has been reconstructed we can  adapt the algorithm  to obtain the posterior distribution for  more complex continuous time multi-state models, like also hidden continuous time or random effect models.  In fact, we are currently investigating the possibility to extend the proposed methodology in order to estimate panel data mixtures of Markov, semi-Markov and time-inhomogeneous Markov models.  These kind of generalizations  are possible both considering finite mixtures with a fixed number of components and also by assuming Dirichlet processes mixtures

The ABC approach proposed by Tancredi \cite{tancredi2019biom} for semi-Markov models,  contrary to our proposed MCMC algorithm,  firstly is only able to  produce an approximation of the posterior distribution  even with an infinite number of iterations being based  on  summary statistics which are not sufficient.  Secondly, increasing the model complexity  by the introduction of additional latent variables without a careful choice of the summary statistics would produce very poor inferential results.   Moreover, the choice of the summary statistics should be model based  and as a consequence a generalization of the ABC approach to more complex models may result not so straightforward. 

Finally note that recently have been developed Monte Carlo methods based on Hamiltonian approximations. These kind of techniques have been efficiently implemented with the software Stan. Anyway, due to the  likelihood intractability it does not seem straightforward  to use directly the Hamiltonian algorithms in the context of 
discretely observed continuous time multi state models.  In fact,  to bypass the likelihood intractability, like in our approach,   it is necessary  to simulate the  trajectories distribution  conditional to the observed  points. Such a trajectories  have  an unknown number of jump points and consequently an unknown number of sojourn times   and require variable dimension proposals for being simulated. 
In our approach via the uniformization algorithm  we propose Markovian  conditional trajectories which are intrinsically variable dimension avoiding to use reversible jump techniques  that represent the standard algorithm for these kind of problems.  Then, to best of our knowledge, it is not possible to directly  use the  Stan implementation within reversible jump proposals.   
}

%We also explored the possibility to perform Bayesian inference for discretely observed continuous time Markov models with transition rates depending on general time-dependent covariates. In fact,  a part from the numerical solutions of the Kolmogorov forward equations devoleped by \cite{titman2011flexible} for this class of models, it is common practice, even in a likelihood-based approach,  to assume piecewise constant  rather than smooth intensity functions.  Then, generalizing our uniformization based Metropolis-Hastings algorithm to reconstruct non-homegeneous Markov trajectories  allows a wider range of time nonhomogeneous Markov models to be fitted to panel data.  %Da "In fact,.." lascio qui, abstract o introduzione?%

%\bibliography{bibliography} 

\newpage

\begin{table}
\begin{center}
\begin{tabular}{| l   c | c  c   c   c  | c c c c|}
\hline
 & & \multicolumn{4}{c|}{$M_1$} & \multicolumn{4}{c|}{$M_2$} \\
\hline
& & $E(\cdot|x)$ & $SD(\cdot|x)$ & $q_{0.025}(\cdot|x)$& $q_{0.0975}(\cdot|x)$ & $E(\cdot|x)$ & $SD(\cdot|x)$ & $q_{0.025}(\cdot|x)$& $q_{0.0975}(\cdot|x)$\\
\hline
$MCMC$& $\gamma_{12}$  & 0.13 & 0.03 & 0.08 & 0.20  & 0.11 & 0.03 & 0.06 & 0.18  \\ 
& $\gamma_{13}$ &&&&&  0.02 & 0.01 & 0.00 & 0.05 \\
&$\gamma_{21}$ & 0.05 & 0.03 & 0.01 & 0.14  & 0.05 & 0.03 & 0.01 & 0.13\\ 
&$\gamma_{23}$   & 0.22 & 0.04 & 0.14 & 0.31   & 0.20 & 0.04 & 0.12 & 0.29 \\
\hline
$ABC$& $\gamma_{12}$  & 0.13 & 0.04 & 0.07 & 0.21 & 0.10 & 0.04 & 0.04 & 0.19\\
& $\gamma_{13}$&&&&  & 0.03 & 0.02 & 0.01 & 0.07 \\
&$\gamma_{21}$ & 0.05 & 0.04 & 0.01 & 0.16 & 0.04 & 0.03 & 0.01 & 0.14 \\
&$\gamma_{23}$ & 0.24 & 0.06 & 0.15 &0.40 &0.19 &0.05 &0.10 &0.32\\
\hline
\end{tabular}

\vspace{0.5 cm}

\begin{tabular}{| l   c | c  c   c   c  | c c c c|}
\hline
 & & \multicolumn{4}{c|}{$M_3$} & \multicolumn{4}{c|}{$M_4$} \\
\hline
& & $E(\cdot|x)$ & $SD(\cdot|x)$ & $q_{0.025}(\cdot|x)$& $q_{0.0975}(\cdot|x)$ & $E(\cdot|x)$ & $SD(\cdot|x)$ & $q_{0.025}(\cdot|x)$& $q_{0.0975}(\cdot|x)$\\
\hline
$MCMC$& $\gamma_{12}$   & 0.16 & 0.06 & 0.08 & 0.32 & 0.14 & 0.06 & 0.06 & 0.30\\
&  $\gamma_{13}$ & & & & & 0.02 & 0.02 & 0.00 & 0.06\\
& $\gamma_{21}$& 0.08 & 0.07 & 0.01 & 0.29 & 0.07 & 0.05& 0.01 & 0.21\\
& $\gamma_{23}$ & 0.34 & 0.16 & 0.16& 0.93 & 0.23 & 0.09 & 0.09 &0.48\\ 
& $\alpha_1$ & 0.83 & 0.19 & 0.49 & 1.22 & 0.82 & 0.19& 0.49 & 1.23 \\
& $\alpha_2$ & 0.71 & 0.17 & 0.43 & 1.11& 0.76 & 0.19 & 0.45 & 1.18 \\
\hline
$ABC$& $\gamma_{12}$  & 0.17 & 0.11 & 0.07 & 0.60 & 0.11 & 0.06 & 0.04 & 0.26\\
& $\gamma_{13}$&&&&  & 0.04 & 0.03 & 0.01 & 0.10 \\
&$\gamma_{21}$ & 0.17 & 0.51 & 0.01 & 0.80 & 0.06 & 0.06 & 0.01 & 0.24 \\
&$\gamma_{23}$ & 0.34 & 0.16 & 0.16 &0.93 &0.23 &0.09 &0.09 &0.48\\
& $\alpha_1$ & 0.83 & 0.19 & 0.49 & 1.22 & 0.82 & 0.19& 0.49 & 1.23 \\
& $\alpha_2$ & 0.72 & 0.19 & 0.44 & 1.15& 0.80 & 0.21 & 0.45 & 1.27 \\

\hline
\end{tabular}

\caption{Breast cancer data set.  
Posterior means, standard deviations and 0.025, 0.0975 quantiles for the the Markov models $M_1$ with $\gamma_{13}=0$
and $M_2$ with $\gamma_{13}>0$ and the corresponding Weibull semi-Markov models $M_3$ and $M_4$. \label{destavola1}   }
\end{center}
\end{table}
 
\clearpage

\begin{table}
\begin{center}

%\begin{tabular}{|l|rrrrrrrrrr}

\begin{tabular}{| p{2.75cm} |  p{0.9cm} p{0.9cm} p{0.9cm} p{0.9cm} p{0.9cm}   p{0.9cm} p{0.9cm}  p{0.9cm} p{0.9cm} p{0.9cm}  |}
  \hline
Markov  & $p_{12}$ & $p_{14}$ & $p_{23}$ &$p_{24}$ &  $\gamma_{1}$&$\gamma_{2}$& $\gamma_{3}$  & & & \\
  \hline
$E(\cdot|x)$ & 0.77 & 0.23 & 0.81 & 0.19 & 0.14 & 0.26 & 0.24   & & &  \\ 
 $SD(\cdot|x)$ & 0.03 & 0.03 & 0.06 & 0.06 & 0.01 & 0.02 & 0.03  & & &  \\ 
 $q_{0.025}(\cdot|x)$  & 0.71 & 0.18 & 0.69 & 0.08 & 0.12 & 0.22 & 0.19   & & &  \\ 
 $q_{0.975}(\cdot|x)$  & 0.82 & 0.29 & 0.91 & 0.30  & 0.15 & 0.30 & 0.29  & & &  \\
  \hline 
\end{tabular}
\vspace{0.3 cm}

\begin{tabular}{| p{2.75cm} |  p{0.9cm} p{0.9cm} p{0.9cm} p{0.9cm} p{0.9cm}   p{0.9cm} p{0.9cm}  p{0.9cm} p{0.9cm} p{0.9cm}  |}
  \hline
Semi Markov & $p_{12}$ & $p_{14}$ & $p_{23}$ &$p_{24}$  &  $\gamma_{1}$&$\gamma_{2}$& $\gamma_{3}$  &$\alpha_1$ & $\alpha_2$ & $\alpha_3$\\
  \hline
$E(\cdot|x)$ & 0.71 & 0.29 & 0.85 & 0.15  & 0.14 & 0.28 & 0.28 & 0.94 & 0.92 & 0.99 \\ 
 $SD(\cdot|x)$  & 0.03 & 0.03 & 0.08 & 0.08 & 0.01 & 0.03 & 0.04  & 0.05 & 0.08 & 0.13 \\ 
 $q_{0.025}(\cdot|x)$   & 0.65 & 0.24 & 0.70 & 0.01 & 0.12 & 0.23 & 0.21  & 0.85 & 0.78 & 0.77 \\ 
 $q_{0.975}(\cdot|x)$  & 0.76 & 0.35 & 0.99 & 0.30  & 0.15 & 0.33 & 0.37 & 1.04 & 1.10 & 1.27 \\
  \hline 
\end{tabular}
\vspace{0.3 cm}

%\begin{tabular}{| p{2.75cm} |  p{0.9cm} p{0.9cm} p{0.9cm} p{0.9cm} p{0.9cm}   p{0.9cm} p{0.9cm}  p{0.9cm} p{0.9cm} p{0.9cm}  |}
%\hline 
%Inohomogeneous  & $p_{12}$ & $p_{14}$ &  $p_{23}$ &$p_{24}$ & $\beta_{0_1}$ & $\beta_{1_1} $ & $\beta_{0_2} $ & $\beta_{1_2} $& $\beta_{0_3} $& $\beta_{1_3} $   \\
%\hline
% $E(\cdot|x)$  & 0.67 & 0.33 & 0.98 & 0.02 & -2.19  & 0.06 & -1.24 & -0.01  & -2.00 & 0.10   \\
% $SD(\cdot|x)$ & 0.03 & 0.03 & 0.02 & 0.02 &\hphantom{.}0.09  & 0.02& \hphantom{.}0.18 & \hphantom{.}0.03  &\hphantom{.}0.23  & 0.02    \\
% $q_{0.025}(\cdot|x)$ & 0.62& 0.27  & 0.93 & 0.00  &-2.36 & 0.02 & -1.59 & -0.07 & -2.47 & 0.05  \\
%$q_{0.975}(\cdot|x)$  & 0.73 & 0.38 & 1.00 & 0.07 &-2.02 & 0.09 & -0.88  & 0.04 & -1.56 & 0.14   \\
%\hline
%\end{tabular}
\begin{tabular}{| p{2.75cm} |  p{0.9cm} p{0.9cm} p{0.9cm} p{0.9cm} p{0.9cm}   p{0.9cm} p{0.9cm}  p{0.9cm} p{0.9cm} p{0.9cm}  |}
\hline 
Inohomogeneous  & $p_{12}$ & $p_{14}$ &  $p_{23}$ &$p_{24}$ & $\beta_{0_1}$ & $\beta_{1_1} $ & $\beta_{0_2} $ & $\beta_{1_2} $& $\beta_{0_3} $& $\beta_{1_3} $   \\
\hline
 $E(\cdot|x)$  & 0.67 & 0.33 & 0.98 & 0.02 & -2.19  & 0.06 & -1.24 & -0.01  & -2.00 & 0.10   \\
 $SD(\cdot|x)$ & 0.03 & 0.03 & 0.02 & 0.02 &\hspace{0.1mm} 0.09  & 0.02& \hspace{0.1mm} 0.18 &  \hspace{0.1mm} 0.03  &\hspace{0.1mm} 0.23  & 0.02    \\
 $q_{0.025}(\cdot|x)$ & 0.62& 0.27  & 0.93 & 0.00  &-2.36 & 0.02 & -1.59 & -0.07 & -2.47 & 0.05  \\
$q_{0.975}(\cdot|x)$  & 0.73 & 0.38 & 1.00 & 0.07 &-2.02 & 0.09 & -0.88  & \hspace{0.1mm} 0.04 & -1.56 & 0.14   \\
\hline
\end{tabular}

\caption{Cardiac allograft vasculopathy dataset.  
Posterior means, standard deviations and 0.025, 0.975 quantiles for the Markov (M), Weibull semi-Markov (SM) and time inhomogeneous Markov (TIM) models with $p_{13}=p_{31}=0$. \label{CAV1}   }
\end{center}
\end{table}

\clearpage

\begin{table}

\begin{center}
\begin{tabular}{|r |c | c|c|c|c| c|c | }
\hline
$n$ & $\gamma_{12}$  &  $\alpha_1$ &  $\gamma_{13} $&   $\gamma_{21}$  &  $\alpha_2$ &  $\gamma_{23}$  \\
\hline
50 &   0.25    (0.06)    &  1.43 (0.30)     & 0.07    (0.02)  &  0.05 (0.03)   &   0.91  (0.20)    &  0.09 (0.03) \\
100 & 0.25   (0.04)      & 1.42 (0.18)    & 0.06 (0.02)   & 0.05 (0.02)   & 0.81 (0.14) & 0.10 (0.03)   \\
500 &0.25 (0.02) &1.39 (0.10)&0.05 (0.01)&0.04 (0.01) & 0.73 (0.07)  & 0.10 (0.01)\\
1000 &      0.25 (0.02)        & 1.40 (0.08)        & 0.05 (0.01)         & 0.04   (0.01)        &  0.72  (0.05)   &  0.10 (0.01) \\
\hline
\end{tabular}

\vspace{0.33 cm}

\begin{tabular}{|r|l | l|l|l|l| l|l | }
\hline
50 &   0.25    (0.05)    &  1.43 (0.23)     & 0.07    (0.02)  &  0.05 (0.03)   &   0.86  (0.17)    &  0.09 (0.03) \\
100 &   0.25 (0.04) & 1.40 (0.17)  & 0.06 (0.02) & 0.05 (0.02)& 0.76 (0.11) & 0.10  (0.03) \\
500 & 0.25 (0.02)& 1.40 (0.07)& 0.05 (0.01)& 0.04 (0.01)& 0.72 (0.06) & 0.10 (0.01) \\
1000 &   0.25  (0.01)    &     1.40 (0.07) &    0.05 (0.01)    &     0.04  (0.01)&    0.71  (0.04)&  0.10 (0.01)\\
\hline
\end{tabular}

\caption{Simulated data: mean and standard deviation (in parentheses) of the posterior means across 100 samples of size $n=(50,100,500,1000)$ under a three states Weibull semi-Markov model with  one absorbing state,  $\theta=(\gamma_{12},\alpha_1,\gamma_{13}, \gamma_{21}, \alpha_2, \gamma_{23})=(0.25,1.4,0.05,0.04,0.7,0.1)$ and follow-up times equal to 0,3,6,12,24,60. Upper table: death time unknown. Lower table: death time exactly known \label{tab_sim_SM}}. .
\end{center}
\end{table}

\clearpage

\begin{table}

\begin{center}
\begin{tabular}{|r |c | c|c|c|c| c| }
\hline
$n$ & $\beta_{0_1}$  &  $\beta_{1_1}$ &  $\beta_{0_2}$ &   $\beta_{1_2}$  \\
\hline
50 &   -0.72 (0.22)   &  0.24 (0.10)   & -2.31(0.39)  &  0.20 (0.06)     \\
100 & -0.70 (0.15)      & 0.21 (0.07)  &-2.31 (0.27)   & 0.20 (0.04)    \\
500 & -0.69 (0.07) & 0.20 (0.03) &-2.29(0.12) & 0.20 (0.02) \\
1000 & -0.69 (0.05) & 0.20 (0.02) &-2.28(0.09) & 0.20 (0.01) \\
\hline
\end{tabular}

\vspace{0.33 cm}

\begin{tabular}{|r |c | c|c|c|c| c|}
\hline
$n$ &  $p_{12}$  & $p_{13}$ & $p_{21}$ & $p_{23}$ \\
\hline
50 &  0.79 (0.06)   &  0.21 (0.06)    & 0.20 (0.09) &  0.80 (0.09)   \\
100 & 0.80 (0.04)  &  0.20 (0.04)  & 0.19 (0.06)  & 0.81 (0.06)  \\
500 & 0.81 (0.02) & 0.19 (0.02) & 0.19 (0.03) & 0.81 (0.03)  \\
1000 & 0.81 (0.02) & 0.19 (0.02) & 0.18 (0.02) & 0.82 (0.02)  \\
\hline
\end{tabular}

\caption{Simulated data: mean and standard deviation (in parentheses) of the posterior means across 100 samples of size $n=(50,100,500,1000)$ under a three states time-inhomogeneous Markov model with absorbing state,  $\theta=(\beta_{0_1},\beta_{1_1},\beta_{0_2}, \beta_{1_2}, p_{12}, p_{13},p_{21},p_{23})=(-0.69,0.2,-2.30,0.2,0.8,0.2,0.2,0.8)$ and follow-up times equal to 0,1,2,3,4.5,6,9,12,20. \label{tab_sim_TIM}}.
\end{center}
\end{table}

\clearpage

\newpage

%\bibliography{wileyNJD-AMA}
\end{document}